# Ultra-sensitive Short-Wave Infrared Single-Photon Detection using a Silicon Single-Electron Transistor

P. Sudha,[1] S. Miyagawa,[2] A. Samanta,[1,3*] and D. Moraru[2*]

[1]Quantum/Nano-Science and Technology Lab, Department of Physics
Indian Institute of Technology Roorkee, Roorkee, Uttarakhand 247667, India
Email: arup.samanta@ph.iitr.ac.in
[2]Research Institute of Electronics, Shizuoka University, 3-5-1 Johoku, Chuo-ku, Hamamatsu 432-8011, Japan
Email: moraru.daniel@shizuoka.ac.jp
[3]Centre of Nanotechnology
Indian Institute of Technology Roorkee, Roorkee, Uttarakhand 247667, India

**Keywords**: Single-photon detector, Short-wave infrared, All silicon technology, Single-electron transistor, Random telegraph signal

Ultra-sensitive short-wave infrared (SWIR) photon detection is a crucial aspect of ongoing research in quantum technology. However, developing such detectors on a CMOS-compatible silicon technological platform has been challenging due to the low absorption coefficient for silicon in the SWIR range. In this study, a codoped silicon-based single-electron transistor (SET) in a silicon-on-insulator field-effect transistor (SOI-FET) configuration is fabricated, which successfully detects single photons in the SWIR range with ultra-high sensitivity. The detection mechanism is evidenced by the shift in the onset of the SET current peaks and by the occurrence of random telegraph signals (RTS) under light irradiation, as compared to the dark condition. The calculated sensitivity of our device, in terms of noise equivalent power, is approximately $10^{-19}$ W Hz$^{-1/2}$.

## 1. Introduction

The detection of single photons in the short-wave infrared (SWIR) range ($\lambda$ = 1000-3000 nm) is critical for a wide range of applications, including spectroscopy[1], quantum information processing[2,3], biomedical imaging[4], communication[5] and sensing[6,7]. Silicon-based single-electron transistor (SET) devices offer a promising approach for single-photon detector by sensing changes in the SET current induced even by individual photons. SETs have previously been utilized for single-photon detection (SPD) in the visible spectrum (for energies larger than the bandgap of silicon, $E_g \approx 1.12$ eV)[7,8] and in the far-infrared spectrum (for energies comparable to the donor-energy ionization energy $\approx 45$ meV in bulk Si).[9] For instance, M. Tabe and collaborators reported SPD in the visible range using a multi-dot array.[7,8] In these devices, the trapping and detrapping of electrons in quantum dots (QDs) near the conduction path under light irradiation modulates the SET current, leading to SPD.

On the other hand, T. Okamoto *et al*. leveraged inter-dopant tunneling and donor-electron detrapping for THz-SPD, attributing the detection capability to the donor-to-donor and donor-to-conduction-band energies being in the THz range.[9] However, Si-based devices face challenges in the SWIR range due to low absorption coefficient ($10^1$ - $10^{-8}$ cm$^{-1}$ in the range of 1000 nm to 1450 nm and negligible for higher wavelength), which correspond to the sub-bandgap region in the case of Si.[10] Current state-of-the-art detectors for this range include avalanche photo-diodes, QD-based systems, and Ge-based SPDs.[11-14] Although these detectors are compatible with the CMOS technology, they do not fully utilize the silicon technology. Additionally, while 2D materials and their heterostructures demonstrate excellent performance for IR detection,[15] their practical implementation remains limited by low absorption due to their ultra-thin layers.[16]

For the utilization of silicon for the sub-bandgap region, higher doping can be employed, since a dopant band can be formed within the bandgap[17]. This effectively reduces the bandgap of doped-Si as compared to the bandgap of intrinsic (non-doped) Si, thereby enabling the detection of relatively larger wavelengths. However, such high doping can result in metal-like behavior, diminishing the probability of single-charge tunneling through the device and, consequently, reducing the SPD sensitivity.

In our approach, we counter-doped boron (B) atoms into the highly phosphorus (P) doped channel to increase the potential barriers between donors. This can help maintain single-charge tunneling even with high donor doping levels.[18,19] In addition, the random distribution of dopants and the nanoscale dimensions of the channel also play crucial roles. The local potential fluctuations induced by codoping in such nanometer-sized channel facilitate single-charge tunneling in the field-effect transistor (FET) configuration.

In this work, we fabricated codoped back-gated silicon-on-insulator (SOI) FETs that exhibit strong SET current characteristics. These devices demonstrate significant detection capabilities of SWIR at $\lambda$ = 1780 nm, which corresponds to the sub-bandgap region of silicon. We observed two distinct SPD mechanisms in our device: (a) the shift of $|I_{DS}|$ along the $V_{BG}$ axis under SWIR irradiation; (b) the trapping and detrapping of photo-generated charges sensed as changes of the SET current ($|I_{DS}|$) in time. Therefore, these devices hold great potential for the development of SWIR SPDs utilizing all-silicon technology.

## 2. Results and Discussion

The device under study is a codoped silicon-on-insulator field-effect transistor (SOI-FET) defined using a CMOS-compatible electron-beam lithography technique with nanometer-scale channel dimensions, as shown in Figure 1(a). The scanning electron microscopy (SEM) image of the device is shown in Figure 1(b). The dimensions of the channel region are approximately 100 nm in length ($l$), 15 nm in width ($w$), and 18 nm in depth ($d$), as presented in Figure 1(c). The device is codoped with phosphorus (P) at a concentration of approximately $N_D \approx 2\times10^{20}$ cm$^{-3}$ and with boron (B) at a concentration of approximately $N_A \approx 5.3\times10^{19}$ cm$^{-3}$, respectively. Figure 1(c) also shows schematically a random distribution of P and B dopants in the channel region. To allow light absorption in the channel region, the formation of a top gate was omitted for this device. Instead, the Si substrate of the SOI wafer, with a low doping concentration of approximately $N_A \approx 10^{16}$ cm$^{-3}$, is used as a back-gate to control the conduction through the channel. All electrical characterization was performed using a probe station in high vacuum at a low temperature ($T$) of 16 K, utilizing a semiconductor precision parameter analyzer (Agilent 4156C).

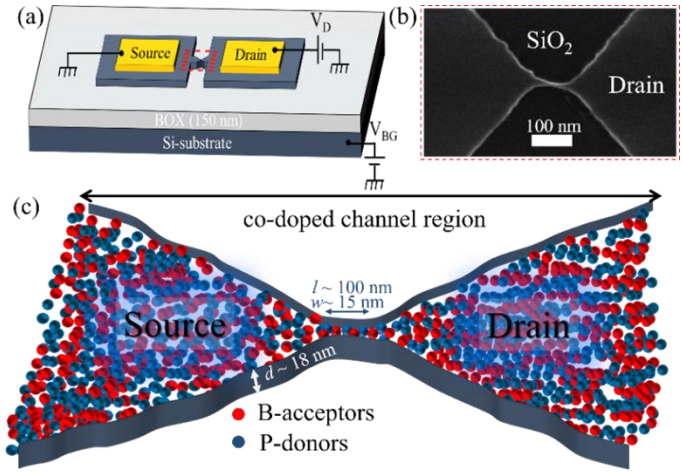

**Figure 1.** (a) Schematic illustration of the codoped SOI-FET with back-gate configuration, with the source grounded during the measurements. The nanoscale channel region is marked by the red dashed square. (b) SEM image of the device in the channel region, with scale bar of 100 nm. (c) Schematic representation of the random distribution of P-donors and B-acceptors within the channel region of the SOI-FET.

The source-to-drain current ($|I_{DS}|$) versus back-gate voltage ($V_{BG}$) characteristics of the device for $V_D$ = 35 mV, 75 mV and 115 mV in the dark are shown in Figure 2(a). First of all, a single, isolated SET current peak was observed at $V_{BG} \approx$ -3.09 V for a large $V_D$ value, likely due to transport via the ground state (GS) of a deep P-donor (labeled $D_1^{+/0}$, where $D_1^{+/0}$ marks the transition from $D^+$ to $D^0$ state for the first donor). As $V_{BG}$ increases, four irregular current peaks appear at $V_{BG} \approx$ -1.848 V, -0.94 V, -0.44 V, and -0.14 V, labeled as $D_2^{+/0}$, $D_3^{+/0}$, $D_4^{+/0}$, and $D_5^{+/0}$, respectively. These peaks likely result from transport through the GS of four different weakly-coupled P-donors, i.e., donors that are close to each other without molecular bonding. The weak coupling between P-donors, despite such high doping, may occur due to the counter-doping with B-acceptors, which may significantly lower the effective doping concentration. It can be noticed that the intensity of these peaks generally decreases from the 2$^{nd}$ to the 5$^{th}$ peak as $V_{BG}$ increases. Normally, increasing the gate voltage would decrease the barrier height and increase the current intensity.[20,21] However, the random location of the donors may have caused variations in tunnel resistances, leading to this unexpected behavior. Another relatively isolated peak appears at $\approx$ 0.52 V, which could correspond to the second charge state of the 1$^{st}$ deep donor (labeled as $D_1^{0/-}$), considering also that this peak is higher in intensity than the previous peak at $\approx$ -3.09 V. In addition, four more SET peaks found at $V_{BG} \approx$ 1.04 V, 1.76 V, 2.10 V, and 2.49 V (labeled as $D_2^{0/-}$, $D_3^{0/-}$, $D_4^{0/-}$, and $D_5^{0/-}$) likely represent the second charge states of the previously mentioned weakly coupled few-donor system. Beyond these peaks, a general increase in current indicates that the conduction band of the channel region reached the Fermi levels of the source and drain.

The stability diagram, which shows $|I_{DS}|$ in the $V_{BG}$-$V_D$ plane, is presented in Figure 2(b). The structure of the Coulomb diamonds observed in this diagram further confirms the Coulomb-blockade transport through a single isolated deep P-donor, followed by transport through a weakly coupled four-donor system. The lever-arm factor ($\alpha = C_G/(C_G+C_S+C_D)$) was calculated for each diamond in the stability diagram, where $C_G$, $C_S$, and $C_D$ are gate capacitance, source capacitance, and drain capacitance, respectively.[22,23] The $\alpha$ values for the donors $D_1$, $D_2$, $D_3$, $D_4$, and $D_5$ are 0.027 ± 0.03, 0.07 ± 0.02, 0.052 ± 0.002, 0.06 ± 0.005, and 0.056 ± 0.003, respectively.

The barrier height, calculated from Arrhenius plots such as shown in the inset of Fig. 2(c), is presented in Fig. 2(c) as a function of $V_{BG}$. The extended linear fit intersects the $V_{BG}$ axis at ~ 2.5 V, which is the condition under which the edge of the conduction band is aligned with source Fermi level.[21] The estimated deepest state is situated at ~95 meV below the edge of the condition band. However, when calculated using the lever-arm factor, the deepest state is 142.5 ± 8.39 meV, which may be an overestimation. Based on the device's transport characteristics (Figures 2(a)-(c)) and the estimated positions of the P-donors (evaluated from the slope analysis presented in Table S1 in the supplementary information), a potential distribution model for the channel region is presented schematically in Figure 2(d).

Lastly, no signatures of hole current could be observed in the transport characteristics of the device, indicating that the B-acceptors are likely fully compensated since $N_D > N_A$. It can thus state that these counter-dopants only assist in increasing the potential barrier between P-donors, thus enabling tunneling effects in the channel.

In order to investigate light-induced effects, the stability diagram was measured under light illumination, as presented in Figure 3(a). The stability diagram reveals a significant shift of $\Delta V_{BG}$ = 2.91 V from $V_{BG}$ = -3.09 V in dark (Figure 2(b)) to $V_{BG}$ = -0.18 V in light (Figure 3(a)) for the first observable current peak. This shift of the SET peaks likely results from a modification of the Fermi levels of the source and drain, which could rise above that of the channel region due to increased dopant activation in the extended source and drain regions under light illumination. This phenomenon is illustrated schematically in Figure 3(b) and detailed explanation is represented schematically in supplementary information, Figure S1. This reconfiguration of the barrier height slightly modifies the transport characteristics of the device, keeping the same number of SET peaks.

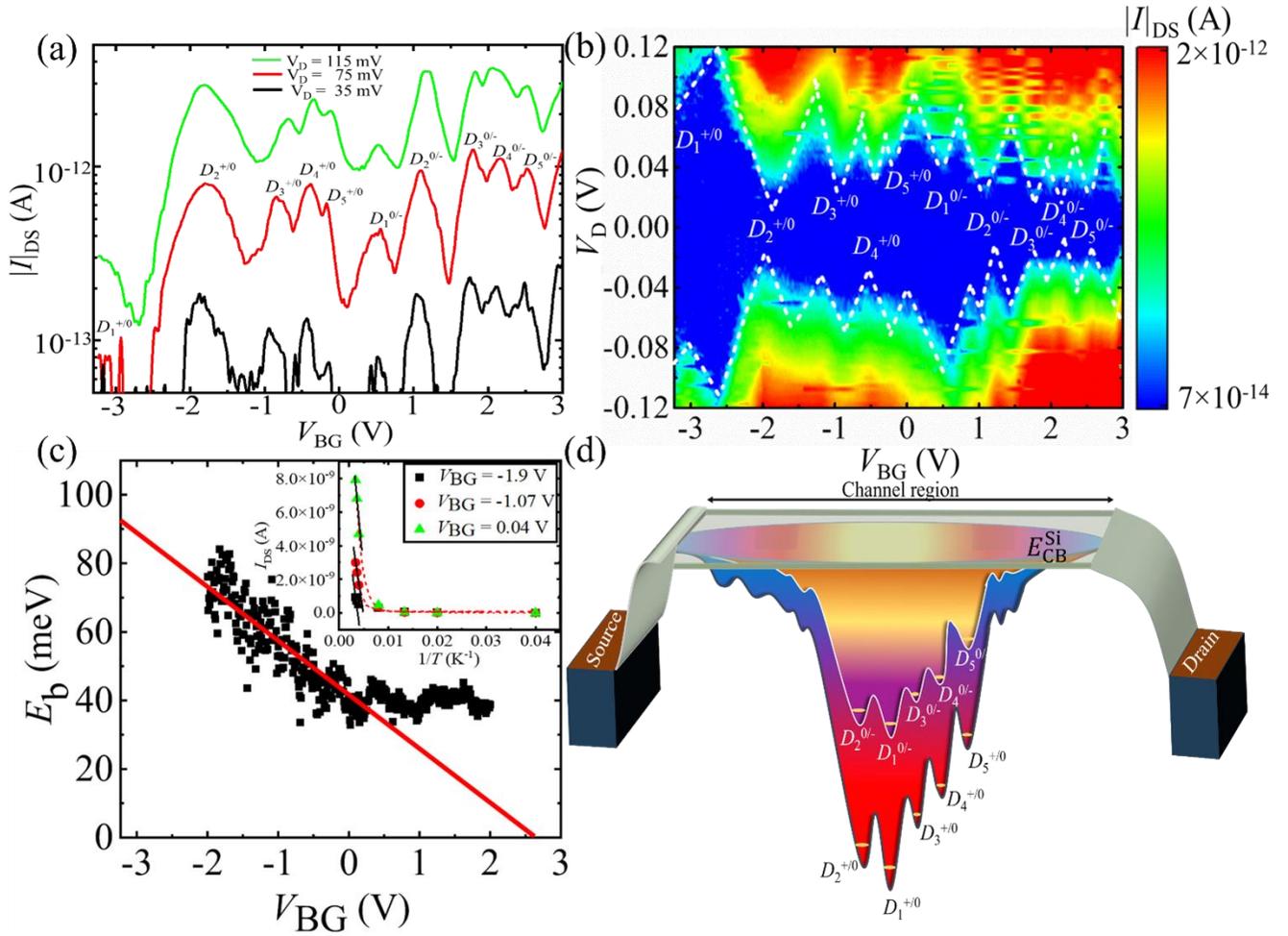

**Figure 2** (a) The $I_D$-$V_{BG}$ characteristics at various $V_D$ values in dark at $T = 16$ K. (b) Stability diagram (plot of $|I_{DS}|$ in the $V_D$-$V_{BG}$ plane), with the Coulomb diamonds highlighted by dashed lines. (c) The deepest energy state is estimated from the Arrhenius plot as a function of $V_{BG}$. The inset shows the Arrhenius plot, for selected $V_{BG}$ values with the corresponding exponential fits. (d) Schematical representation of the potential landscape in the channel region, illustrating the transport mechanism through the donor-induced potential wells. $E_{CB}^{Si}$ is the conduction band edge of silicon.

For observing individual photo-induced effects, current $|I_{DS}|$ was measured versus time in dark and in SWIR ($\lambda = 1780$ nm) light-illumination conditions at $V_{BG} = 0.1$ V and $V_D = 0.08$ V, as shown in Figure 3(c). Considering the observed shift, the dark current is taken including the estimated shift of the onset characteristics. In the dark, the current $|I_{DS}|$ remains almost constant, as shown in Figure 3(c), bottom panel. Under light illumination, however, random telegraph signals (RTS) are clearly observed, as shown in Figure 3(c), upper several panels. The occurrence probability of the dark or light current levels in the RTS is further analyzed by weighted time-lag plots (w-TLP), [24] presented on the right side of each panel of Figure 3(c). The w-TLP method is an extension of the time-lag plot (TLP) analysis, where the RTS's $j^{th}$ points are plotted on the x-axis and its $(j+1)^{th}$ points are plotted on the y-axis, representing the entire RTS trace. The weight is added in each point of TLP by the difference in position between different $(x, y)$ points, optimizing w-TLP to enhance the current level of RTS while minimizing the background noise. This w-TLP analysis shows a single current level in the dark, as seen in the right side of Figure 3(c), bottom panel. Under light illumination at $\lambda = 1780$ nm with a flux density

$\phi \approx 0.30$ nm$^{-2}$s$^{-1}$, two distinct levels of the RTS were observed, with current levels $I_1$ ($\approx 1.5\times10^{-13}$ A) and $I_2$ ($\approx 7\times10^{-13}$ A). The w-TLP analysis of the RTS data confirms statistically the existence of these two current levels. Notably, $I_1$ corresponds to the dark current level, whereas $I_2$ is additionally observed under light illumination. This pronounced current shift from $I_1$ to $I_2$ suggests detrapping from a trap site near the transport path, with the trap being most probably a donor in this system. As the flux density increased to $\phi \approx 0.37$ nm$^{-2}$s$^{-1}$ and $\phi \approx 0.42$ nm$^{-2}$s$^{-1}$, multiple RTS levels were incorporated in the measured current. The corresponding w-TLP analysis indicates the existence of 3 and, respectively, 5 currents levels for $\phi \approx 0.37$ nm$^{-2}$s$^{-1}$ and $\phi \approx 0.42$ nm$^{-2}$s$^{-1}$. Together with the $I_1$ and $I_2$ current levels, a strong current level $I_3 \approx 4\times10^{-13}$ is also introduced for both flux densities $\phi \approx 0.37$ nm$^{-2}$s$^{-1}$ and $\phi \approx 0.42$ nm$^{-2}$s$^{-1}$, as clearly seen in the corresponding w-TLP plots. This level likely arises after the trapping of a detrapped electron into another nearby trap site. In addition, even more trap sites participate in transport at higher flux density, $\phi \approx 0.42$ nm$^{-2}$s$^{-1}$. Since only the deepest level is filled, it is reasonable to consider that detrapping is also induced by single photons.

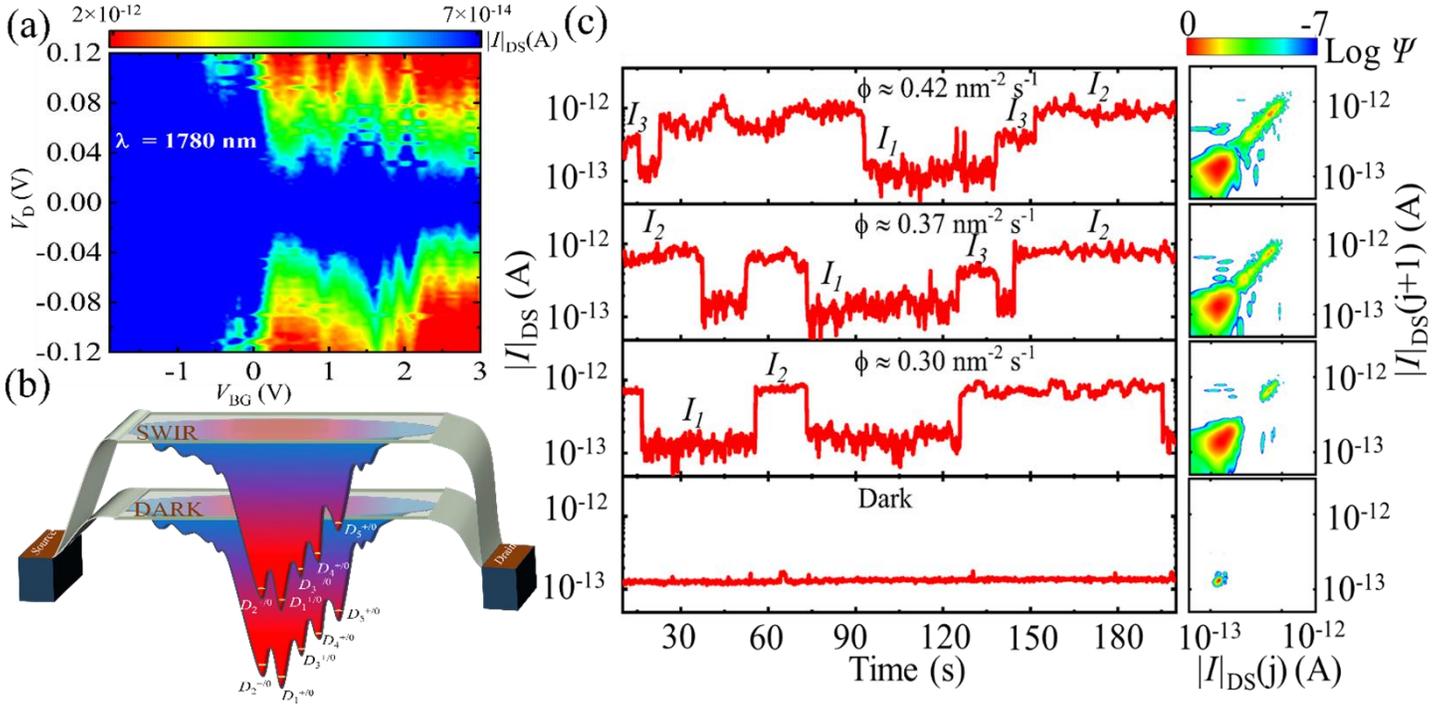

**Figure 3.** (a) Stability diagram under light illumination with a wavelength $\lambda = 1780$ nm. (b) Schematic representation of the mechanism responsible for the observed shift in the stability diagram under light (SWIR) illumination. (c) Current $|I_{DS}|$ as a function of time at $V_{BG} = 0.1$ V and $V_D = 80$ mV for SWIR illumination ($\lambda = 1780$ nm). Random telegraph signals (RTS) are observed for various intensities of SWIR light, further analyzed by the w-TLP method, as shown in the right-most panels. The RTS current levels are marked in Figure 3(c). The measurement time-step is $\Delta t = 200$ ms.

Moreover, the stability diagrams under different photon energies were also measured, for $\lambda = 1154$ nm (corresponding to energies involving excitation of carriers between valence band and the donor band near the conduction band edge) and $\lambda = 750$ nm (condition for which the energy is larger than the bandgap), as shown in Figures 4(a) and 4(b), respectively. We observed comparable shifts in the transfer characteristics along $V_{BG}$ for both $\lambda = 1154$ nm and $\lambda = 750$ nm. Since the flux density is similar for these stability diagrams, the average number of activated dopants under light illumination is likely the same, leading to the nearly equal shifts observed in our device. The number of trapping and detrapping events also increases as the wavelength decreases, which is evident even from the stability diagrams. Following this, the current $|I_{DS}|$ was measured versus time for $\lambda = 1154$ nm and $\lambda = 750$ nm, with different photon flux shown in the supplementary information presented in Figures S2. Various events of trapping and detrapping can be noticed in these plots.

For comparing the effects of shorter wavelengths with $\lambda = 1780$ nm, the $|I_{DS}|$ – Time (t) plots are shown for nearly identical $\phi$ values in Figure 4(c). New current levels are more prominently observed in RTS for these smaller wavelengths, as presented by the w-TLP plots on the right side of each panel of Figure 4(c). This may occur because new paths for charge transitions are now accessible, from valence band to the conduction band or to the donor-band. From the w-TLP plots presented earlier, two current levels were identified for $\lambda = 1780$ nm, but three levels were found for $\lambda = 1154$ nm and five levels for $\lambda = 750$ nm. We also calculated the number of RTS events per second, $N_{RTS} = 1/(\tau_{emp} + \tau_{occ})$, where $\tau_{emp}$ and $\tau_{occ}$ are the average times of the empty and occupied trap states, respectively. We obtained $N_{RTS}$ values of 0.153s$^{-1}$, 0.069s$^{-1}$, and 0.064 s$^{-1}$ for $\lambda$ of 750 nm, 1154 nm, and 1780 nm, respectively. The large $N_{RTS}$ observed for relatively large wavelength in our device is attributed most likely to the higher doping concentration. Based on the results shown so far, it can be expected that both mechanisms, the change in SET current under light irradiation compared to dark, as shown by the stability diagrams, and the RTS generation with a few distinct levels, can be adapted for SPD.

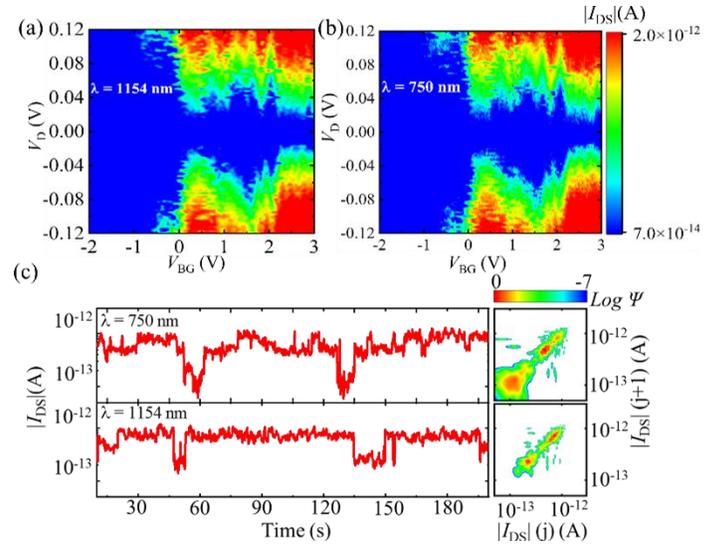

**Figure 4.** Stability diagrams under light illumination for different wavelengths: (a) $\lambda = 1154$ nm and (b) $\lambda = 750$ nm. (c) Current $|I_{DS}|$ plotted as a function of time at $V_{BG} = 0.1$ V and $V_D = 80$ mV for $\lambda = 1154$ nm and $\lambda = 750$ nm. The measurement time-step is $\Delta t = 200$ ms.

The detector sensitivity is estimated based on the dark switching rate ($t_{ds}^{-1}$) and the quantum efficiency ($\eta$). Since the current level generally remains constant in the dark, the dark switching rate is negligible. We assumed $t_{ds}^{-1} \approx 1000$ s$^{-1}$. The quantum efficiency is the ratio of detected photons to the incident photons. Thus, we calculated $\eta$ by determining the reflectance and absorption coefficients. [25]

To determine the absorption coefficient of SOI, with thickness similar to that of the channel of the device under study, we conducted spectroscopic ellipsometry measurements across a wavelength range from 250 nm to 1688 nm. As expected, the measured absorption coefficient is increased in comparison with the bulk-Si wafer. We extracted the absorption coefficient for the device layer, and the corresponding data for the absorption coefficient and the absorption depth as a function of wavelength are provided in the supplementary information (Figure S3(a) and (b), respectively). Giving the increasing trend of the absorption coefficient measured up to $\lambda = 1688$ nm from $\lambda = 250$ nm and owing to the fact that both 1688 nm and 1780 nm wavelengths correspond to the sub-bandgap region, we infer that the increasing trend likely extends to $\lambda = 1780$ nm. From the ellipsometry data, we determined the reflectance and absorption coefficients. [26] Accordingly, the estimated detector sensitivity for the SWIR wavelength in terms of noise equivalent power (NEP) [24] was calculated as $NEP = \frac{\sqrt{2\pi}\hbar\omega}{\eta\sqrt{t}} = 5.89 \times 10^{-19}$ W Hz$^{-1/2}$. By comparison, the current state-of-the-art sensitivity using a Ge-on-Si device at $\lambda = 1310$ nm is $1.9 \times 10^{-16}$ W Hz$^{-1/2}$ at $T = 78$ K. [13] This means that, remarkably, we report in this work an even higher sensitivity at longer wavelengths using all-silicon technology.

## 3. Conclusion

In this study, we have demonstrated a highly sensitive all-silicon single-photon detector operating in the short-wave infrared (SWIR) wavelength region, utilizing a codoped silicon-on-insulator field-effect transistor (SOI-FET) as a single-electron transistors (SET). The fundamental method used for the photon detection in the SWIR region relies on the analysis of the random telegraph signals (RTS) and of the relative shift of the SET peaks under SWIR illumination as compared to dark conditions. Moreover, the high doping concentration in our device effectively enhances photon absorption, extending the capabilities of silicon into the sub-bandgap SWIR region. Notably, our device also shows potential for detecting single photons in the near infrared (NIR) and visible ranges. The sensitivity estimated for SWIR photons, with a noise equivalent power (NEP) of approximately $10^{-19}$ W Hz$^{-1/2}$, represents a significant advancement compared to existing technologies. This ultra-high sensitivity, achieved through an SET-based detection mechanism, positions our system as a promising candidate for future applications in long-wavelength photon detection.

## Experimental section

### Device fabrication

The device was fabricated using silicon-on-insulator (SOI) wafer with a buried oxide (BOX) layer of 150 nm thickness. The fabrication processes employed CMOS-compatible techniques in clean room environment. The initial SOI thickness of 55 nm was reduced to a final thickness of $t_{Si} \approx 18$ nm by sacrificial oxidations and standard oxidation processes. Electron-beam lithography (EBL) was used for all patterning processes, including channel patterning, with precise control over the channel width. The top oxide layer was formed by thermal dry oxidation at 800°C, 15 minutes for obtaining a thickness of approximately 10 nm as a final gate oxide. Aluminum (Al) was used for the gate, source, and drain contacts.

### Electrical measurements

Electrical measurements were performed using mainly an Agilent 4156C precision semiconductor parameter analyzer with a variable-temperature probe station. All current-voltage (I-V) measurements are carried out in high vacuum. The light illumination setup, which included an optical fiber, a monochromator, a voltage source and a halogen lamp, was integrated with the prober for accurate measurements under various light illumination conditions.

## Acknowledgments

We appreciate valuable discussions with Prof. M. Tabe on this line of research. We also acknowledge S. Chakraborty for fruitful discussions.

## Data Availability Statement

All the data available within the manuscript and supplementary information.

## Funding Statement

P. Sudha acknowledges the Council of Scientific and Industrial Research (CSIR) for fellowship. This work is supported by project STARS-2/2023-0715 funded by IISc-MHRD, India and SR/FST/PS-II/2019/84 funded by DST, India.

## Conflict of Interest

The authors declare no conflict of interest.

# Supporting Information

## Ultra-sensitive Short-Wave Infrared Single-Photon Detection using a Silicon Single-Electron Transistor

P. Sudha,[1] S. Miyagawa,[2] A. Samanta,[1,3*] and D. Moraru[2*]

[1]*Quantum/Nano-Science and Technology Lab, Department of Physics*
*Indian Institute of Technology Roorkee, Roorkee, Uttarakhand 247667, India*
Email: arup.samanta@ph.iitr.ac.in
[2]*Research Institute of Electronics, Shizuoka University, 3-5-1 Johoku, Chuo-ku, Hamamatsu 432-8011, Japan*
Email: moraru.daniel@shizuoka.ac.jp
[3]*Centre of Nanotechnology*
*Indian Institute of Technology Roorkee, Roorkee, Uttarakhand 247667, India*

## Location of Dopants:

The positions of dopants (P-donors) within the channel were determined by analyzing the slopes of each Coulomb diamond from the stability diagram in Fig. 2. The slope analysis results are provided in Table S1. For the case of a grounded-source single-electron transistor (SET) circuit, the positive and negative slopes of the Coulomb diamonds flanking a certain current peak are $C_G / (C_G + C_S)$ and $-C_G / C_D$, respectively. The ratio of these slopes provides an estimate of the position of the quantum dot, here the P-donor, responsible for the SET peak along the channel length, expressed in terms of the effective channel length ($L'$).

**Table S1.** Positions of P-donors estimated relative to the source, expressed as a fraction of the effective channel length ($L'$) for our device.

| Dopant | Position |
|--------|----------|
| $D_1$  | $0.52L'$ |
| $D_2$  | $0.46L'$ |
| $D_3$  | $0.58L'$ |
| $D_4$  | $0.62L'$ |
| $D_5$  | $0.67L'$ |

## Detailed schematical representation of the significant shift in stability diagram under SWIR:

Figure S1a represents the average electron distribution in the device and Figure S1 (b) represents the corresponding Fermi level of the device in dark without any source-drain bias ($V_{DS}$).

After illuminating the device with SWIR the average number of electron concentration is increased as presented in Figure S1 (c) which can increase the Fermi level of the whole system. However, the channel size is relatively small compared to source and drain, thus the change in Fermi level will also be comparable to the dimensions as shown in Figure S1 (d), such shift in Fermi level under SWIR compared to dark is shown by black arrow in Figure S1 (d). Moreover, as soon as SWIR illuminates the device the Fermi level will shift and realign as presented in Figures from S1 (d) to (e). This process will be so fast. Thus, this new Fermi level will increase the conduction band energy, which induces the current level shifting under light.

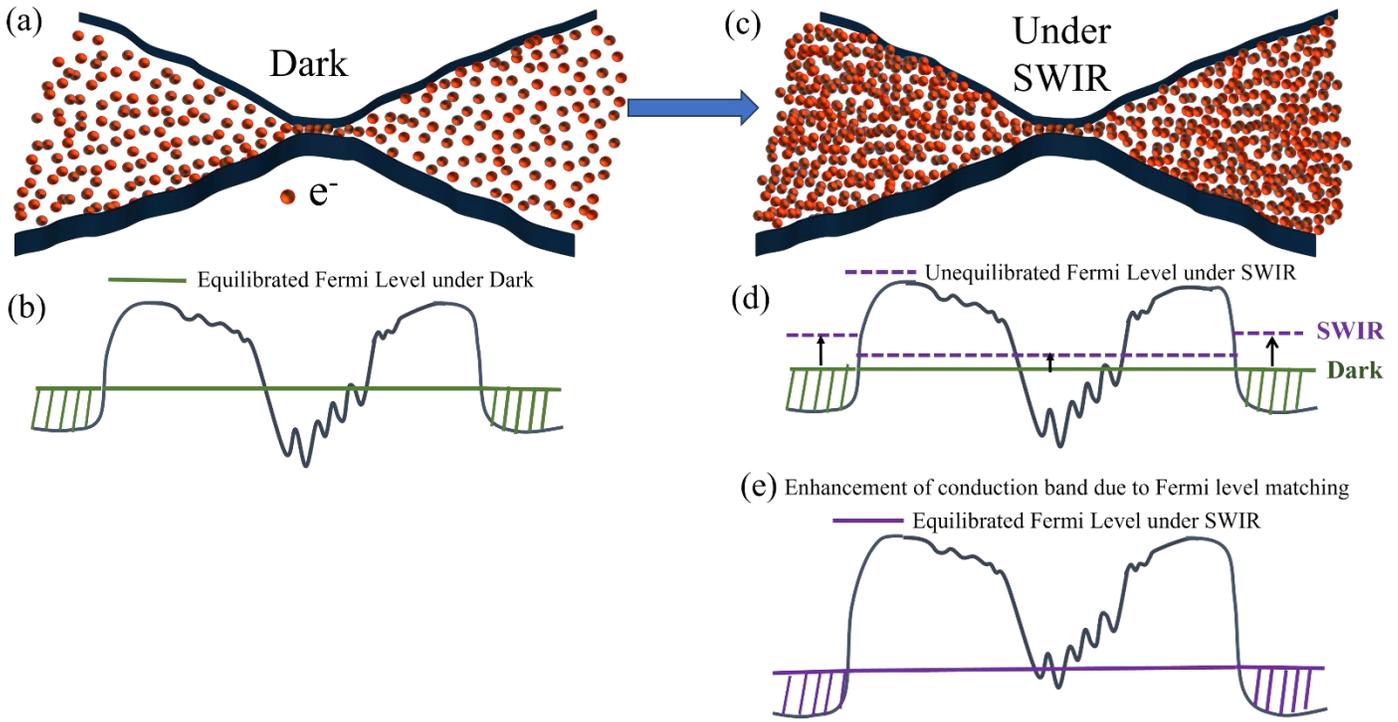

**Figure S1.** (a) Electron distribution in the device in dark. (b) The corresponding Fermi level in dark. (c) The electron distribution under SWIR illumination. (d) The corresponding unequilibrated Fermi level under SWIR. The black arrow shows the shift in Fermi level under SWIR from dark' Fermi level. (e) The corresponding equilibrated Fermi level under SWIR.

## RTS measurement for $\lambda = 1154$ nm and $\lambda = 750$ nm:

Additional measurements were conducted for the current $|I_{DS}|$ versus time in dark and light conditions, for different wavelengths: $\lambda = 1154$ nm and $\lambda = 750$ nm, as shown in Figures S1 and S2. For $\lambda = 1154$ nm, a noticeable change in the dark current level was observed, indicating a shift to a new state. The trapping and detrapping events varied for different wavelengths, likely due to the higher energy of the photons, which may introduce new channels for charge transitions.

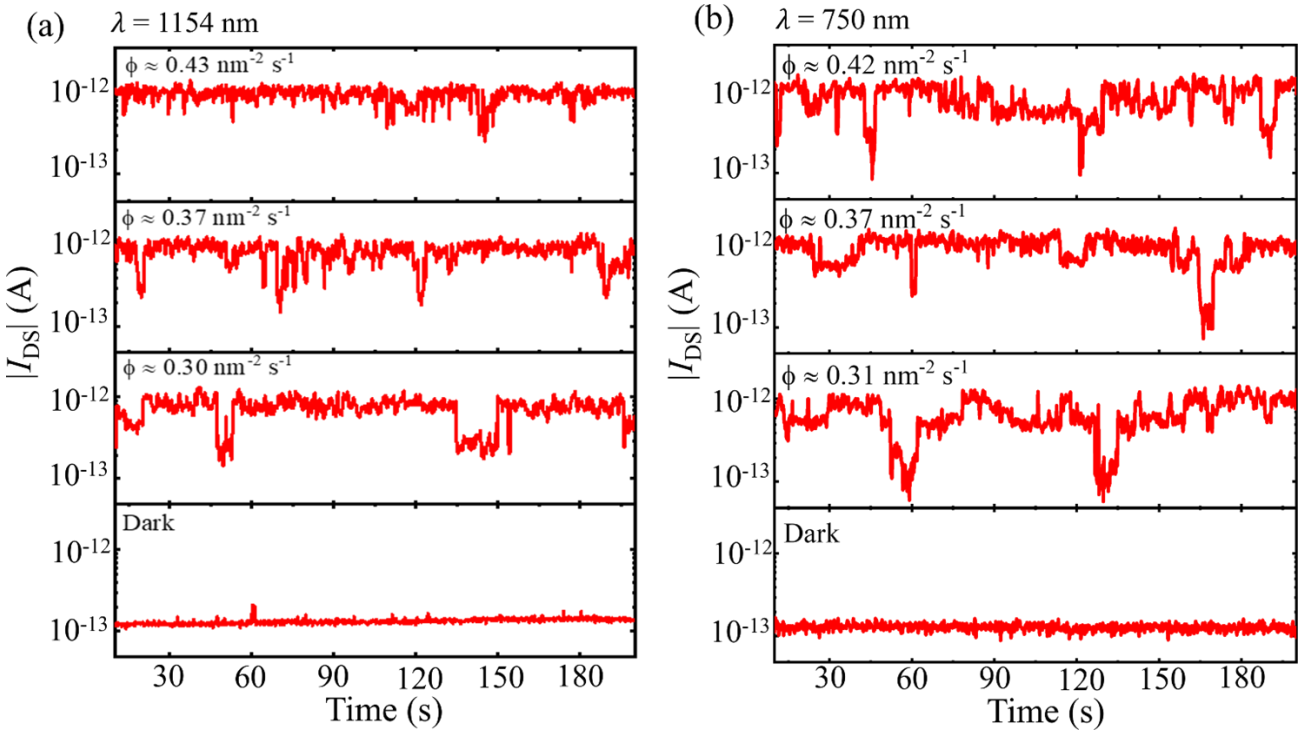

**Figure S2.** Current $|I_{DS}|$ as a function of time at $V_{BG} = 0.1$ V and $V_D = 80$ mV for: (a) for $\lambda = 1154$ nm and (b) $\lambda = 750$ nm. The RTS behavior is shown for various light intensities at both wavelengths. The measurement time-step is $\Delta t = 200$ ms.

## Absorption coefficient of the device layer:

To estimate the sensitivity of the device in terms of noise equivalent power (NEP), we extracted the absorption coefficient for the top-silicon layer from a silicon-on-insulator (SOI) sample with doping concentrations comparable to the main device under study. We also measured the absorption coefficient of the substrate silicon (bare-silicon) extracted from the data obtained for the SOI wafer and that of a bare-silicon wafer using Spectroscopic Ellipsometer. Data are shown in Figure S2. The measurements demonstrate an increasing trend in the absorption coefficient from $\lambda = 250$ nm to $\lambda = 1688$ nm. Considering that both 1688 nm and 1780 nm correspond to energies in the sub-bandgap region, we can infer that this increasing trend likely extends to $\lambda = 1780$ nm, as well.

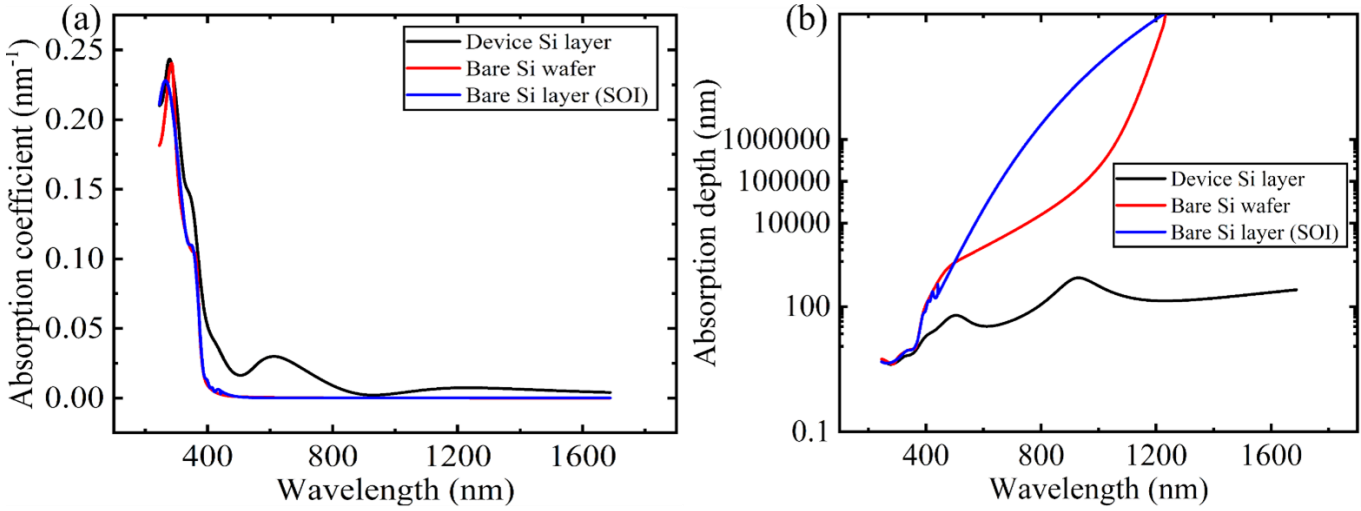

**Figure S3.** Spectroscopic ellipsometer measurements of: (a) the absorption coefficient and (b) the absorption depth as a function of wavelength.